\pdfoutput=1
\documentclass[aps,pre,superscriptaddress,amsmath,amssymb,reprint,nofootinbib,twoside]{revtex4-2}
\usepackage{graphicx}
\usepackage{bm}

\begin{document}

\title{Using Lagrangian descriptors to calculate the Maslov index of periodic orbits}
\author{J. Montes}
\email{jmontes.3@alumni.unav.es}
\affiliation{Grupo de Sistemas Complejos, Universidad Polit\'ecnica de Madrid, 28040 Madrid, Spain}
\affiliation{Departamento de Qu\'\i mica, Universidad Aut\'onoma de Madrid, 28049 Madrid, Spain}
\author{F. J. Arranz}
\email{fj.arranz@upm.es (corresponding author)}
\affiliation{Grupo de Sistemas Complejos, Universidad Polit\'ecnica de Madrid, 28040 Madrid, Spain}
\author{F. Borondo}
\email{f.borondo@uam.es}
\affiliation{Departamento de Qu\'\i mica, Universidad Aut\'onoma de Madrid, 28049 Madrid, Spain}
\date{\today}

\begin{abstract}
The Maslov index of a periodic orbit is an important piece in the semiclassical quantization of non-integrable systems, while almost all existing techniques that lead to a rigorous calculation of this index are elaborate and mathematically demanding. In this paper, we describe a straightforward technique, for systems with two degrees of freedom, based on the Lagrangian descriptors. Our method is illustrated by applying it to the two-dimensional coupled quartic oscillator.
\end{abstract}

\maketitle

\section{\label{sec:intro}Introduction}

At present the most important theoretical result concerning the semiclassical quantization of chaotic systems remains perhaps the celebrated Gutzwiller trace formula~\cite{Gutzwiller.trace.formula}. In this formula, the quantized density of states, which peaks sharply at each quantized energy, is expressed as a sum over the periodic orbits (POs) of the system, where the classical action in Planck constant units $S/\hbar$ appears as a phase factor. Moreover, the reflections of the PO with the potential energy boundaries bring a phase loss that, in the case of an unstable PO, is given by $\mu\frac{\pi}{2}$, where $\mu$ is the so-called Maslov index of the PO. Consequently, the calculation of these indices is a principal step in the procedure of quantization of Gutzwiller. For a more general scope of the Maslov index see, e.g., Ref.~\cite{Beck.Maslov.index}.

\vspace{2ex}   

Notice that, although the development of standard quantum mechanics has led to more efficient methods, semiclassical quantization still remains a topic of interest, mainly due to the fact that it allows us to establish a correspondence between classical and quantum mechanics. In the last few years, many papers have been published where the semiclassical quantization, and more specifically the quantization of POs, is used. Thus, the quantization of POs of Gutzwiller has been used to obtain an optimal basis set for the representation of eigenstates~\cite{Quartic.Maslov.calc.winding.number}, to calculate the level density of the H\'enon-Heiles potential in the context of the symmetry breaking problem~\cite{guztwiller_POs_1}, has been extended for steady states of nonadiabatic systems~\cite{guztwiller_POs_2}, to calculate different properties in nuclear collective dynamics~\cite{guztwiller_POs_3}, to obtain the exceptional points in the elliptical three-disk scatterer~\cite{guztwiller_POs_4}, to achieve an alternative (continuous time) representation of quantum maps on the torus~\cite{guztwiller_POs_5}, to propose an alternative approach to the periodic orbit theory of spectral correlations~\cite{guztwiller_POs_6}, to unveil the creation mechanism of Devil's Staircase surface in the anisotropic Kepler problem~\cite{guztwiller_POs_7}, to study quantum chaos in many-body systems~\cite{guztwiller_POs_8}, to develop a chaotic lattice field theory in one dimension~\cite{guztwiller_POs_9}, and also to obtain the frontier of scars determining the order-chaos transition when the Planck constant is continuously changed~\cite{Arranz.LiCN.POquantization.frontier}.

\vspace{2ex}   

It is noteworthy that in Gutzwiller original derivation~\cite{Gutzwiller.trace.formula}, the parameter $\mu$ was not identified as a Maslov index, but rather as the number of conjugate points (i.e., the points where caustics occur) over one period of the PO. Later, Creagh {\em et al.}~\cite{Creagh.Maslov.calc.winding.number} showed that $\mu$ is an intrinsic property of the PO, since it is independent of the coordinate system used. Specifically, these authors demonstrated that the parameter $\mu$ is determined by a winding number, namely, it is equal to twice the number of times the stable and unstable manifolds rotate around the PO over one period. Then, the parameter $\mu$ is a topological invariant of the PO and can be properly called {\em Maslov index} in the usual sense. Also, Creagh {\em et al.}~\cite{Creagh.Maslov.calc.winding.number} showed that, in general, the Maslov index of a PO cannot be identified with the number of caustics, since this number is not a topological invariant of the PO which can depend on the set of coordinates used and the starting point of the orbit.

\vspace{2ex}   

Moreover, the rigorous calculation of the Maslov index of a PO is not a simple task. There are some easy but non rigorous methods, such as the count of the turning points in each degree of freedom over one period, which give the correct result in many cases but fail in others. There are also rigorous but elaborate and mathematically demanding methods. Some time ago, Eckhardt and Wintgen~\cite{E-W.Maslov.calc.winding.number} developed the theoretical result of Creagh {\em et al.}~\cite{Creagh.Maslov.calc.winding.number} obtaining computationally useful formulas, which are based on the linearized classical mechanics, to calculate the number of half-turns that the invariant manifolds rotate around the PO. Recently, Vergel {\em et al.}~\cite{Maslov.calc.geometrodynamics} showed how the conjugate points of a PO can be computed by using the geometrodynamic approach to classical dynamics, the number of them being equal to the value of the Maslov index. Both methods are rigorous, albeit laborious and not easy to implement. Now, in his paper, we propose a rigorous and straightforward method to obtain, for systems with two degrees of freedom, the number of half-turns that the invariant manifolds rotate around the PO, hence the Maslov index, which is based on the calculation of the Lagrangian descriptors (LDs) along the PO.

In any case, it is worth mentioning the development carried out by Robbins, one of the authors of Ref.~\cite{Creagh.Maslov.calc.winding.number}, who extended to arbitrary dimensions the results achieved in that reference, and also obtained a rigorous yet simple method to calculate the Maslov index as a winding number in arbitrary dimensions (see Ref.~\cite{Robbins.Maslov.calc.winding.number} for a condensed derivation of the method, and references therein for additional details).

\vspace{1ex}   

The LDs are a fruitful mathematical tool to investigate dynamical systems that has given rise to many applications in the last 15 years. This success is largely due to the fact that the calculation is simple and straightforward, producing a faithful graphical representation of the invariant structures existing in phase space. The LDs ``based on the arc length measure'' were first introduced heuristically by Madrid and Mancho~\cite{Mancho.LDs.first_definition} in the context of the theory of aperiodic dynamical systems. Almost immediately this tool was used in geophysical problems concerning the study of ocean flows~\cite{Mancho.LDs.ocean_flows,Mancho.LDs.ocean_flows_Kuroshio,Mancho.LDs.ocean_flows_predictive,Mancho.LDs.ocean_flows_flightMH370,Mancho.LDs.ocean_flows_oil,Mancho.LDs.ocean_flows_North_Atlantic} and also atmospheric dynamics~\cite{LDs.atmospheric_dynamics_RWB,Mancho.LDs.atmospheric_dynamics_warming_I,Mancho.LDs.atmospheric_dynamics_warming_II}. Later the LDs have been used in chemical physics problems regarding the transition state theory~\cite{LDs.chem_phys.TST_1,LDs.chem_phys.TST_2,LDs.chem_phys.TST_3,LDs.chem_phys.TST_4,LDs.chem_phys.TST_5,LDs.chem_phys.TST_6,LDs.chem_phys.TST_7}, isomerization dynamics~\cite{LDs.chem_phys.isomer_dyn_1,LDs.chem_phys.isomer_dyn_2,LDs.chem_phys.isomer_dyn_3,LDs.chem_phys.isomer_dyn_4} (including the reactive islands issue~\cite{LDs.chem_phys.reac_isl_1,LDs.chem_phys.reac_isl_2,LDs.chem_phys.reac_isl_3}), dynamical matching~\cite{LDs.chem_phys.dyn_match_1,LDs.chem_phys.dyn_match_2,LDs.chem_phys.dyn_match_3,LDs.chem_phys.dyn_match_4}, and other interesting cases on chemical reaction dynamics. This tool has also been used to characterize regular motion in dynamical systems~\cite{LDs.Montes_1} and to analyze chaotic billiards, such as the Bunimovich stadium~\cite{LDs.Montes_2}. Recently the LDs have also been applied to biomedical problems (cardiovascular flows~\cite{LDs.biomed_1}, circadian rhythms~\cite{LDs.biomed_2}) and space science (asteroid dynamics~\cite{LDs.space_sci_1}, ballistic capture~\cite{LDs.space_sci_2}). Moreover, the original definition based on the arc length measure was extended by Mancho and co-workers by considering different measures~\cite{Mancho.LDs.different_definitions} and also discrete-time dynamical systems, i.e., maps~\cite{Mancho.LDs.maps}. Also, a response to some criticisms~\cite{Ruiz-Herrera.LDs.criticism_1,Ruiz-Herrera.LDs.criticism_2,Haller.LDs.criticism} was given in Ref.~\cite{Mancho.LDs.theoretical_proofs}, where the heuristic methodology of the previous papers was turned to rigorous mathematical proofs and a general definition based on the \mbox{$p$-norm} measure was established.

\vspace{1ex}   

The outline of the paper is as follows. Section~\ref{sec:system.calculations} is devoted to the description of the Hamiltonian system used to illustrate our method (Sec.~\ref {sec:system}), as well as the calculations to obtain the LDs (Sec.~\ref{sec:LDs.calculations}) and eventually the Maslov index (Sec.~\ref{sec:Maslov.calculations}). Section~\ref{sec:results.discussion} is devoted to the joint presentation and discussion of the results obtained by applying our method to some selected POs of the Hamiltonian system used. Last, in Sec.~\ref{sec:conclusions}, the paper is summarized and the main conclusions are given.

\section{\label{sec:system.calculations}System description and calculations}

\subsection{\label{sec:system}Hamiltonian system}

We will illustrate the proposed method by applying it to the two-dimensional coupled quartic oscillator with unit mass defined by the Hamiltonian function
\begin{equation}
\label{eq:Hamiltonian}
H = \frac{1}{2}\left( p_x^2 + p_y^2 \right) + \frac{1}{2} x^2 y^2 +
\frac{\beta}{4}\left( x^4 + y^4 \right),
\end{equation}
where $(x,y)$ are Cartesian coordinates, $(p_x,p_y)$ their corresponding conjugate momenta, and $\beta\ge0$ is a chaos parameter. Note that the condition $\beta\ge0$ guarantees that the system remains bounded. This system has been extensively used in connection with the topic of quantum chaos, and is known to be highly chaotic for most values of the chaos parameter and integrable only for $\beta=1/3$ and $\beta=1$. In our case we have taken $\beta=1/100$, which is a typical value used to study the quartic oscillator in a highly chaotic regime. In particular, this value was used in Refs.~\cite{Quartic.Maslov.calc.winding.number} and~\cite{Maslov.calc.geometrodynamics}, where the Maslov index of several POs is rigorously calculated by counting the winding number and by using the geometrodynamic method, respectively, such that we can check our results against the values of the Maslov index obtained in these references.

Moreover, due to the fact that the energy potential $V=x^2y^2/2+\beta(x^4+y^4)/4$ is a homogeneous function [i.e., $V(ax,ay)=a^nV(x,y)$, with homogeneity order $n=4$], the system exhibits mechanical similarity. This means that any trajectory $\biglb(x(t),y(t),p_x(t),p_y(t)\biglb)$, with energy $E$, scales to another one $\biglb(x^\prime(t^\prime),y^\prime(t^\prime),p_x^\prime(t^\prime),p_y^\prime(t^\prime)\biglb)$, with energy $E^\prime$, such that
\begin{equation}
\label{eq:homogeneous.scaling}
\begin{pmatrix}
x^\prime(t^\prime)   \\
y^\prime(t^\prime)   \\
p_x^\prime(t^\prime) \\
p_y^\prime(t^\prime) \\
\end{pmatrix}
=
\begin{pmatrix}
\gamma & 0 & 0 & 0   \\
0 & \gamma & 0 & 0   \\
0 & 0 & \gamma^2 & 0 \\
0 & 0 & 0 & \gamma^2 \\
\end{pmatrix}
\begin{pmatrix}
x(t)   \\
y(t)   \\
p_x(t) \\
p_y(t) \\
\end{pmatrix}
,
\end{equation}
with $t^\prime=t\gamma^{-1}$, $\gamma=(E^\prime/E)^{1/n}$, and $n=4$ being the order of homogeneity. The mechanical similarity involves that the phase space structure is the same for all energy values, it simply scales with energy. Thus, for simplicity, we have taken the value $E=1$ in our calculations, which is the same value used in Refs.~\cite{Quartic.Maslov.calc.winding.number} and~\cite{Maslov.calc.geometrodynamics} mentioned above.

\subsection{\label{sec:LDs.calculations}Lagrangian descriptors}

In order to comply with the mathematical rigor achieved in Ref.~\cite{Mancho.LDs.theoretical_proofs}, above mentioned in Sec.~\ref{sec:intro}, we have used the general definition for the LDs based on the \mbox{$p$-norm} measure established there. In this way, for a Hamiltonian system with $N$ degrees of freedom, the LDs $M$ are defined as follows
\begin{equation}
\label{eq:LDs.definition}
M_\pm({\bm\zeta}_0;\alpha,\tau) = \pm\sum_{k=1}^{2N}\int_0^{\pm\tau}|\dot{\zeta}_k(t)|^\alpha\,\text{d}t,
\end{equation}
where ${\bm\zeta}=(\zeta_1,\ldots,\zeta_{2N})$ is the vector formed by the $N$ position variables and their corresponding $N$ conjugate momenta, i.e., in our case ${\bm\zeta}(t)$ is the solution of Hamilton equations with initial condition ${\bm\zeta}_0=(\zeta_{1_0},\ldots,\zeta_{2N_0})$, at time $t=0$. Observe that LDs are a function depending on the initial condition ${\bm\zeta}_0$ and two fixed parameters, the exponent $\alpha\in(0,1]$ and the integration time $\tau\in(0,+\infty)$. Notice that, in the case of an unstable PO, backward $M_{-}$ and forward $M_{+}$ forms in Eq.~(\ref{eq:LDs.definition}) lead to obtaining the unstable and stable invariant manifolds, respectively. The overall LDs $M$, as commonly used in the literature, are given by the sum of both forms, namely, $M=M_{-}+M_{+}$.

For the two-dimensional coupled quartic oscillator in Eq.~(\ref{eq:Hamiltonian}), we have $N=2$ and ${\bm\zeta}=(x,y,p_x,p_y)$. Additionally, we have taken the value $\alpha=1$ for the exponent, which corresponds to the integration of the so-called {\em taxicab} norm~\cite{Krauser.taxicab.geometry} of the Hamiltonian flow $\dot{\bm\zeta}(t)$ in Eq.~(\ref{eq:LDs.definition}), and the value $\tau=5.55\,\lambda^{-1}$ for the integration time, where $\lambda$ is the stability exponent of the corresponding PO. Note that the choice of these values is heuristic, i.e., it is necessary to prove with different guesses until obtaining the clearest depiction of the invariant manifolds.

In order to calculate the Maslov index of a PO, different initial conditions ${\bm\zeta}_0=(x_0,y_0,p_{x_0},p_{y_0})$ must be taken along the PO in configuration space, exploring the energetically accessible momentum space at each position, as described in the next subsection.

\subsection{\label{sec:Maslov.calculations}Maslov index}

The proposed method to calculate the Maslov index of a PO is based on the calculation of the number of half-turns that the invariant manifolds rotate around the PO, while this calculation is performed graphically by means of the LDs calculated on a suitable surface of section along the PO. Since the LDs produce a graphical representation of the invariant manifolds, the PO itself being part of these invariant manifolds, by using the surface of section described below the number of half-turns that the invariant manifolds rotate around the PO will be correspond to the number of times the line representing an invariant manifold and the line representing the PO cross.

Thus, for a given total energy (recall that $E=1$, in our case), we define a surface of section along the PO in configuration space by parameterizing the position coordinates $(x,y)$ on the PO by means of the normalized length of the path, $Q$, such that $Q=0$ corresponds to the minimum value of the $x$ coordinate of the PO, $Q=0.5$ corresponds to the maximum value, and $Q=1$ again corresponds to the minimum value of the $x$ coordinate. Notice that the choice of the origin $Q=0$ is not significant. We have chosen the origin described above in order to obtain a representation where $Q=0,1$ and $Q=0.5$ correspond to the left and right, respectively, simultaneous turning points in the case of the selected POs with simultaneous turning points used to illustrate the method. Note that simultaneous turning points are those points of the trajectory where all momentum values vanish at the same time value. Moreover, at each position $Q$ in configuration space, all energetically accessible momentum values will be explored by parameterizing the momentum coordinates $(p_x,p_y)$ by means of the form
\begin{equation}
\label{eq:P.LDs}
P = \left( \phi - \phi^\text{PO} \right) \| {\bf p} \|,
\end{equation}
where $\| {\bf p} \|$ and $\phi\in\left[\phi^\text{PO}-\pi,\phi^\text{PO}+\pi\right]$ rad are the modulus and angle, respectively, of the vector ${\bf p}=(p_x,p_y)$ in momentum space, $\phi^\text{PO}$ being the angle corresponding to the PO for the given position $Q$.

In this way, for a specified PO, the initial condition ${\bm\zeta}_0=(x_0,y_0,p_{x_0},p_{y_0})$ of the LDs, $M_\pm({\bm\zeta}_0)$, is given by the parameterized position coordinates $(x_0,y_0)=(x,y)_{Q}$ and the parameterized momentum coordinates $(p_{x_0},p_{y_0})=(p_x,p_y)_{Q,P}$, such that the initial condition ${\bm\zeta}_0$, and hence the LDs $M_\pm({\bm\zeta}_0)$, will be a function of the parameterizing coordinates $(Q,P)$, namely $M_\pm(Q,P)$. Last, in order to avoid a double counting of the number of half-turns around the PO of the associated invariant manifolds, only one of the two (either the stable or the unstable) invariant manifolds is used by taking either forward $M_+(Q,P)$ or backward $M_-(Q,P)$ form.

Notice that, in the LDs of a PO calculated in this way, the PO itself corresponds to the line $P=0$. Consequently, the number of half-turns around the PO of the represented (stable or unstable) invariant manifold will be given by the number of times the curve representing the invariant manifold crosses the line $P=0$.

\begin{figure}[t]
\includegraphics{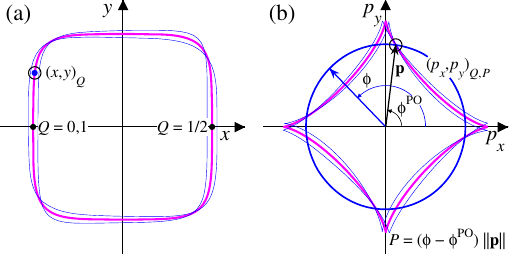}
\caption{\label{fig:PQ_variables}%
Representation in position (a) and momentum (b) space of the parameterization variables $(P,Q)$ defining the surface of section used in the Maslov index calculation. The periodic orbit and two nearby trajectories belonging to an invariant manifold are represented by thick magenta (light) and thin blue (dark) lines, respectively. The open circle ($\bigcirc$) represents an illustrative point $(x,y,p_x,p_y)$ of the periodic orbit. The blue (dark) dot ($\bullet$) in position space (a) and the blue (dark) circular curve in momentum space (b) represent, respectively, the parameterization $(x,y)_Q$ and $(p_x,p_y)_{P,Q}$. Notice that the illustrative point of the periodic orbit corresponds to a caustic in momentum space.}
\end{figure}
\begin{figure}[t]
\includegraphics{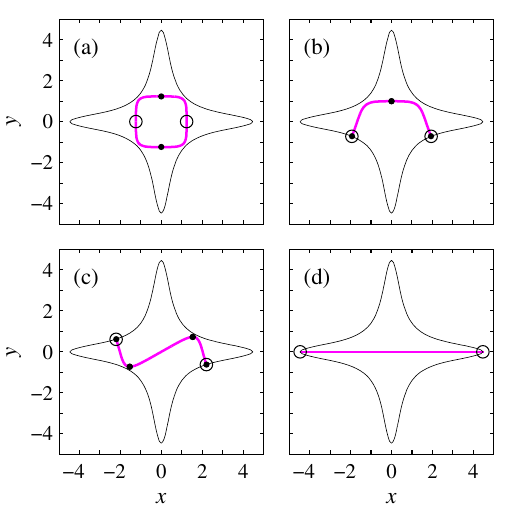}
\caption{\label{fig:PO_turning_points}%
Configuration space representation of the selected periodic orbits, which are referred to in the text as PO-A (a), PO-B (b), PO-C (c), and PO-D (d). The turning points in the $x$ coordinate are marked with open circles ($\bigcirc$) while the turning points in the $y$ coordinate are marked with dots ($\bullet$). The graph of each periodic orbit and the corresponding energy contour are represented by thick magenta and thin black lines, respectively.}
\end{figure}
The parameterization variables $(P,Q)$ defining the surface of section are represented in Fig.~\ref{fig:PQ_variables}. On the one hand, we depict in Fig.~\ref{fig:PQ_variables}~(a) the position space representation of a PO, including two nearby trajectories belonging to an invariant manifold. Notice that these two trajectories are not closed, i.e., they are not POs, as it seems in the scale of the figure, but slowly converge (diverge) to (away from) the central PO. Observe that, as defined above, $Q=0$ corresponds to the minimum value of the $x$ coordinate of the PO, $Q=0.5$ corresponds to the maximum value, and $Q=1$ again corresponds to the minimum value of the $x$ coordinate, closing the PO. Thus, for a given value $Q$, we have the corresponding point $(x,y)_Q$ in position space. On the other hand, we depict in Fig.~\ref{fig:PQ_variables}~(b) the momentum space representation of the PO, also including both nearby trajectories belonging to the invariant manifold. Observe that, as described above in Eq.~(\ref{eq:P.LDs}), for the given point $(x,y)_Q$ in position space, we have the corresponding point in momentum space marked by the vector ${\bf p}$ in the figure. Then, the rotation $\phi$ of the vector ${\bf p}$ describes the circular curve depicted in the figure, which represents all energetically accessible momentum values $(p_x,p_y)_{P,Q}$ for the given point $(x,y)_Q$ in position space. Notice that the given point of the PO selected in Fig.~\ref{fig:PQ_variables} just corresponds to a caustic in momentum space, i.e., a half-turn of the invariant manifold around the PO. Also, considering the definition of the variables $(P,Q)$, the corresponding LDs $M_\pm(Q,P)$ will lead to a crossing of the curve representing the invariant manifold with the line $P=0$ at the given point $Q$.

Moreover, as pointed out by Eckhardt and Wintgen~\cite{E-W.Maslov.calc.winding.number} for winding numbers calculated in configuration space rather than in the whole phase space, in the case of self-retracing POs (i.e., POs with simultaneous turning points) a value of 2, corresponding to both simultaneous turning points, must be added in order to obtain the correct Maslov index.

\section{\label{sec:results.discussion}Results and discussion}

\begin{figure}[t]
\includegraphics{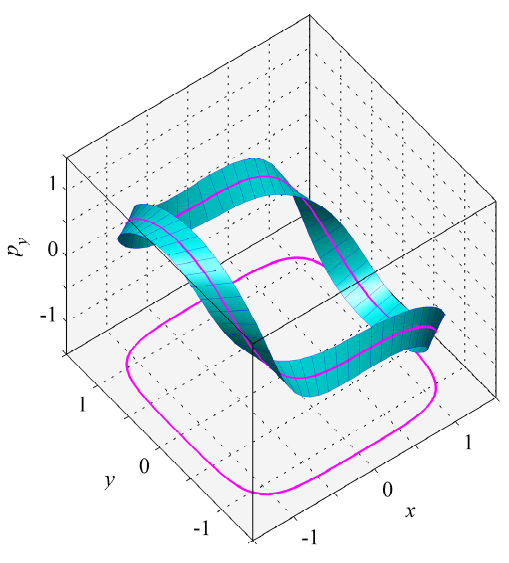}
\caption{\label{fig:linear_manifold}%
Linear stable invariant manifold corresponding to the periodic orbit depicted in Fig.~\ref{fig:PO_turning_points}~(a). Note that the actual linear width around the periodic orbit is greatly increased. The periodic orbit is represented by a thick magenta line.}
\end{figure}
\begin{figure*}[t]
\includegraphics{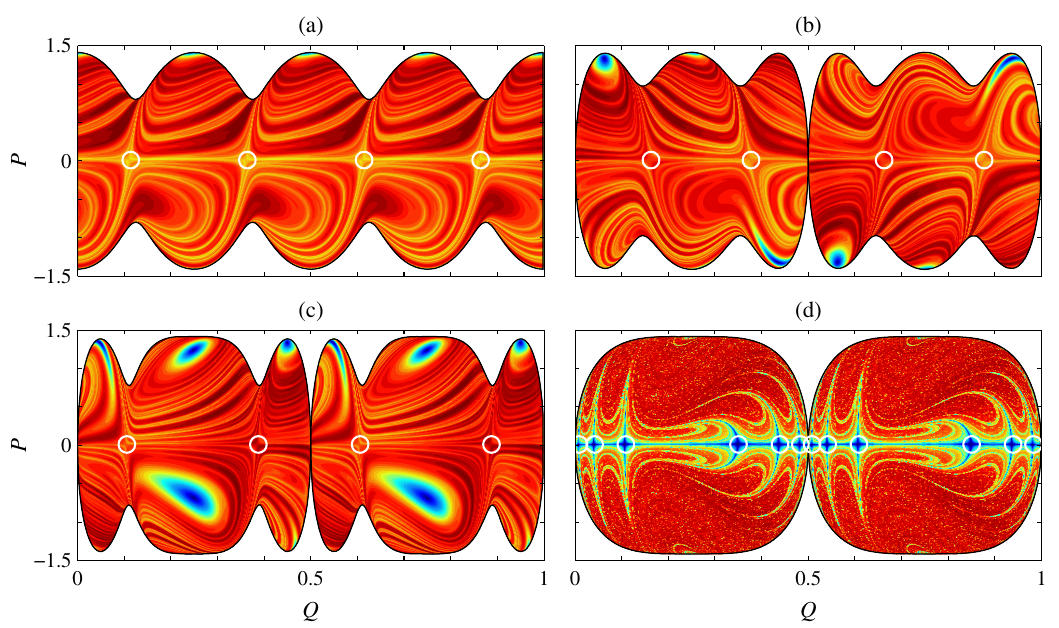}
\caption{\label{fig:LD_maslov}%
Color scale representation of the forward form $M_+(Q,P)$ of the Lagrangian descriptors calculated along the selected periodic orbits depicted in Fig.~\ref{fig:PO_turning_points} over one period. Colorless white area represents the energetically inaccessible region. The crossings of the stable invariant manifold with the line $P=0$ are marked with white open circles ($\bigcirc$).}
\end{figure*}
We have selected four different POs of the quartic oscillator in Eq.~(\ref{eq:Hamiltonian}) in order to illustrate the application of the proposed method: the PO-A without simultaneous turning points in $(x,y)$ coordinates (a closed O-like graph, in the configuration space), the PO-B with both simultaneous turning points on the same side (an open C-like graph), the PO-C with both simultaneous turning points on opposite sides (an open S-like graph), and the PO-D which is simply a one-dimensional straight line (an open I-like graph). The configuration space representation of these POs is depicted in Fig.~\ref{fig:PO_turning_points}, where the turning points in each coordinate have been marked.

\begin{table}[b]
\caption{\label{tab:POs.parameters}%
Values of the significant parameters related to the selected periodic orbits depicted in Fig.~\ref{fig:PO_turning_points}. For each periodic orbit, the period $T$, the stability exponent $\lambda$, the integration time $\tau$, and the Maslov index $\mu$ are listed.}
\begin{ruledtabular}
\begin{tabular}{ccccc@{ }}
Periodic orbit & $T$ & $\lambda$ & $\tau$ & $\mu$ \\
\hline
PO-A &\ 7.8432 & 0.7120 &\ 7.7949 &\ 4 \\
PO-B & 10.6899 & 0.6469 &\ 8.5794 &\ 6 \\
PO-C & 11.8709 & 0.6237 &\ 8.8985 &\ 6 \\
PO-D & 16.5837 & 0.1014 & 54.7337 & 16 \\
\end{tabular}
\end{ruledtabular}
\end{table}
Additionally, the values of the significant parameters related to these POs (period, stability exponent, integration time for LDs calculation, and Maslov index) are given in Table~\ref{tab:POs.parameters}. The values of the Maslov indices given in Table~\ref{tab:POs.parameters} correspond to those obtained in Refs.~\cite{Quartic.Maslov.calc.winding.number} and~\cite{Maslov.calc.geometrodynamics}, where the Maslov indices were calculated using different rigorous methods, and they also correspond to the Maslov indices obtained using our proposed method, as will be shown in this section. Note that, as indicated in Sec.~\ref{sec:LDs.calculations} for the calculation of the LDs, the ratio of integration time to inverse of the stability exponent is maintained at $\tau/\lambda^{-1}= 5.55$ in all cases.

First, it is illustrative to apply the easy, but non rigorous, method to obtain the Maslov index consisting in the count of the turning points in each degree of freedom over one period. We observe in Fig.~\ref{fig:PO_turning_points} that, in the case of the PO-A, the number of turning points over one period is 4 (2 in the $x$ coordinate and 2 in the $y$ coordinate), so that it matches the true Maslov index. For the PO-B, the number of turning points over one period is 6 (2 in the $x$ coordinate and 4 in the $y$ coordinate), so that it also matches the true Maslov index. However, for the PO-C the number of turning points over one period is 8 (2 in the $x$ coordinate and 6 in the $y$ coordinate), which does not match the true Maslov index being 6. Last, the case of the PO-D shows a spectacular discrepancy with the true value. Although it is an ostensibly very simple PO, namely, a one-dimensional straight line in the $x$ coordinate, and hence with 2 turning points over one period (2 in the $x$ coordinate and 0 in the $y$ coordinate), the true Maslov index however is 16, so that the matching definitely fails by far in this case.

Moreover, since the proposed method is based on the graphical calculation of the number of half-turns that the invariant manifolds rotate around the PO, it can be enlightening to depict an example of the behavior of an invariant manifold in the vicinity of the PO (i.e., in the linear regime where the results of Eckhardt and Wintgen~\cite{E-W.Maslov.calc.winding.number} were obtained). Thus, in Fig.~\ref{fig:linear_manifold} we have represented the stable invariant manifold, in the linear regime, corresponding to the PO-A. Note that, in order to obtain a clearer graphical representation, the actual width of the invariant structure in the linear regime has been greatly increased. We observe in this figure how the strip representing the linear invariant manifold is twisted twice, so that it rotates 4 half-turns around the PO, and indeed the Maslov index of the PO-A is 4.

However, as shown below, we can obtain the number of half-turns that the invariant manifolds rotate around the PO in a much simpler way by using the LDs. The forward form $M_+(Q,P)$ of the LDs calculated along each of the four selected POs (PO-A, PO-B, PO-C and PO-D) over one period is depicted in Fig.~\ref{fig:LD_maslov}. On the one hand, in the four cases we observe throughout the range of the $Q$ coordinate a horizontal line at $P=0$, which corresponds to the PO itself. Indeed, this line is the locus were stable and unstable invariant surfaces intersect at the origin of the tangent space of the PO. Consequently, this horizontal line must evidently appear in the LDs calculated on the surface of section described in Sec.~\ref{sec:Maslov.calculations}. On the other hand, in the four cases we also observe a series of ostensibly different lines crossing the line $P=0$, which are approximately straight in the vicinity of the crossings while they stretch and twist as recede from these ones. These crossing lines correspond to the two branches of the stable invariant manifold, hence they are not different lines but a single line which is given by the intersection of the stable invariant manifold and the surface of section defined in Sec.~\ref{sec:Maslov.calculations}. The aforementioned stretching and twisting, resulting from the chaotic character of the quartic oscillator system, hinder the visualization of this geometric object as a single line. Note that, since the value $P=0$ corresponds to the momentum of the PO, at each crossing point the corresponding branch of the invariant manifold coincides with the PO, i.e., it determines a turn of the branch around the PO. Therefore, by counting the successive crossing points we are counting the alternating turns of each branch, namely, we are counting the number of half-turns that the stable invariant manifold rotates around the PO over one period.

\begin{figure}[t]
\includegraphics{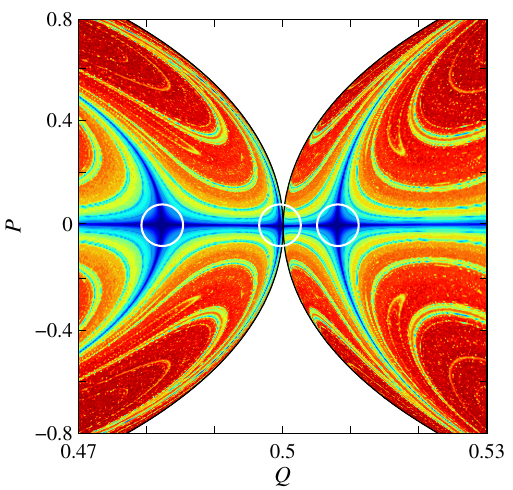}
\caption{\label{fig:LD_maslov_magnification}%
Magnification of Fig.~\ref{fig:LD_maslov}~(d) around the turning point at $Q=0.5$. Observe, just before the turning point, the existence of an additional crossing of the stable invariant manifold with the line $P=0$.}
\end{figure}
In this way, denoting the number of crossing points of the invariant manifold with the line $P=0$ by $N_\text{cp}$, and the number of simultaneous turning points%
\footnote{Note that, regarding the configuration space representation, it holds $N_\text{tp}=0$ for POs with closed graphs and $N_\text{tp}=2$ for POs with open graphs.}
by $N_\text{tp}$, the Maslov index is obtained as $\mu=N_\text{cp}+N_\text{tp}$ for the selected POs. For the PO-A, as shown in Fig.~\ref{fig:LD_maslov}~(a), we obtain $N_\text{cp}=4$ and, since the graph of PO-A is closed, we have $N_\text{tp}=0$, hence its Maslov index is $\mu=4+0=4$. For the PO-B, we obtain from Fig.~\ref{fig:LD_maslov}~(b) the value $N_\text{cp}=4$ and, the graph of PO-B being open, we have $N_\text{tp}=2$, resulting in a Maslov index $\mu=4+2=6$. The case of the PO-C, although some characteristics as the number of turning points in each degree of freedom are not the same, is similar to the PO-B. Thus, for the PO-C we obtain from Fig.~\ref{fig:LD_maslov}~(c) the value $N_\text{cp}=4$ and, the graph of PO-C being also open, we have $N_\text{tp}=2$, resulting in the same Maslov index $\mu=4+2=6$. Last, the case of the PO-D is the most interesting one, since in this case a PO with a very simple graph leads to a high value of the Maslov index. Additionally, the PO-D presents a crossing just before each turning point, which can go unnoticed if an adequate resolution is not used in the depiction of the LDs. Thus, we would obtain from Fig.~\ref{fig:LD_maslov}~(d) the wrong value $N_\text{cp}=12$ for the number of crossings. However, following the behavior of the different crossing lines, we observe a last line which seems to cross the line $P=0$ just before each turning point. Indeed, the magnification depicted in Fig.~\ref{fig:LD_maslov_magnification} shows how this last line actually crosses the line $P=0$ just before the turning point at $Q=0.5$. Obviously, due to the symmetry of the LDs, a magnification around the value $Q=1$ will give an identical figure. Consequently, for the PO-D we obtain the value $N_\text{cp}=14$ and, the graph of PO-D being open, we have $N_\text{tp}=2$, resulting in the Maslov index $\mu=14+2=16$.

Notice that the values obtained by means of the proposed method for the Maslov index of the selected POs are the same as the values obtained in Ref.~\cite{Quartic.Maslov.calc.winding.number}, where the Maslov index was calculated by directly counting the winding number, and also in Ref.~\cite{Maslov.calc.geometrodynamics}, where the Maslov index was calculated by using the geometrodynamic method.

Last, it should be noted that although we have applied the proposed method to a system in a highly chaotic regime, it is applicable without changes to systems in a mixed chaos regime, where regular and chaotic regions coexists in the phase space. Indeed, the method focus on the immediate vicinity of the PO, i.e., the tangent space where the linear approximation is available, and consequently the surrounding regions beyond the tangent space does not affect our results.

\section{\label{sec:conclusions}Summary and conclusions}

The Maslov index of a PO is an important piece in the semiclassical quantization of non-integrable systems developed by Gutzwiller, since it determines the phase loss corresponding to the PO in the Gutzwiller trace formula~\cite{Gutzwiller.trace.formula}. However, almost all existing techniques leading to a rigorous calculation of these indices are elaborate and mathematically demanding. That being the case, we have developed an easy but rigorous and straightforward technique based on the calculation along the corresponding PO of the LDs, which are a fruitful mathematical tool to investigate dynamical systems that have given rise to many applications in the last years.

On the one hand, the proposed method to calculate the Maslov index of a PO is based on the calculation of the number of half-turns that the invariant manifolds rotate around the PO, and therefore it is a rigorous method supported on the findings of Creagh {\em et al.}~\cite{Creagh.Maslov.calc.winding.number}, above mentioned in Sec.~\ref{sec:intro}. On the other hand, the calculation of the number of half-turns is graphically performed by means of the depiction of the LDs calculated on a suitable surface of section along the PO, and therefore the method is straightforward as is the calculation of the LDs.

Additionally, the proposed method has been applied to four different POs of the two-dimensional coupled quartic oscillator, and the obtained results have been positively checked against the values obtained in the literature using different rigorous techniques~\cite{Quartic.Maslov.calc.winding.number,Maslov.calc.geometrodynamics}.

Concerning limitations, we would note that our method is restricted to systems with two degrees of freedom, such that an extension to higher dimensions would be a desirable further development. However, the way to achieve this extension is not clear.

On the one hand, the Maslov index of a PO is determined by a winding number, namely, the number of times the invariant manifolds turn around the PO over one period. But in the general case these turns can be negative or positive (clockwise or anti-clockwise), such that a negative turn cancels a positive one in the total count over one period. The method we propose in principle only detects each half-turn when the invariant manifold crosses the line $P=0$, the direction of rotation remaining undefined. However, we think that a change in the direction of rotation could induce a recognizable pattern in the graphical representation of the LDs, such that identifying this pattern the correct winding number would be obtained.

On the other hand, for higher dimensions, we cannot {\em a priori} reduce the momentum space to only one parameter $P$ in order to define the surface of section $(Q,P)$ for the calculation of the LDs. To address this problem, we think that the systematic study of the LDs calculated on the different surfaces of section $(Q,p_k)$ for each momentum coordinate $p_k$ could shed light on the problem, although the resulting method may not be as simple as in the case of two degrees of freedom. A different way would be to found, if it exists, a distinguished parameter ${\cal P}$ which characterizes the momentum space, perhaps as some kind of average, such that the LDs calculated on the corresponding surface of section $(Q,{\cal P})$ give the winding number in the same fashion as in the two degrees-of-freedom case. In this way the resulting method would be as simple as the method proposed in this paper for two degrees of freedom.

\begin{acknowledgments}
This research was supported by the Spanish Ministry of Science and Innovation under contract No.\ PID2021-122711NB-C21 ({\sc ChaSisCOMA} project).
\end{acknowledgments}

\end{document}